\documentclass[preprint,aps,eqsecnum,nofootinbib]{revtex4}
\usepackage{epsfig}
\usepackage{graphicx}

\begin{document}
\begin{flushright}
HIP-2004-10/TH\\
\end{flushright}

\title{Curvatons in the minimally supersymmetric standard model}

\author{Kari Enqvist}

\affiliation{
Department of Physical Sciences, University of Helsinki\\
and
\\ Helsinki Institute of Physics,\\
     P. O. Box 64, FIN-00014 University of Helsinki, Finland.}

%

\begin{abstract}
Curvaton is an effectively massless field whose energy density during inflation
is negligible but which later becomes dominant. This is a novel mechanism to generate
the scale invariant perturbations. I discuss the possibility that the curvaton
could be found among the fields of the minimally supersymmetric standard model (MSSM),
which contains a number of flat directions along which the renormalizable
potential vanishes. The requirements of late domination and the absence of
damping of the perturbations pick out essentially a unique candidate for the
MSSM curvaton. One must also require that inflation takes place in a hidden sector.
If the inflaton energy density can be radiated into extra dimensions, many constraints
can be relaxed, and the simplest flat direction consisting of the Higgses $H_u$ and
$H_d$ would provide a working example of an MSSM curvaton.
\end{abstract}


\maketitle

\section{Introduction}
The recent WMAP observations \cite{WMAP} of the temperature
fluctuations of the cosmic microwave background provide convincing
evidence for a period of cosmic inflation in the early Universe.
The observed perturbation spectrum is nearly scale invariant and
the total density $\Omega_{\rm tot}=1$, as predicted by many
models of inflation (for a review, see ~\cite{Riotto}). These are
usually based on the slow rolling of a scalar field (the
inflaton) in a flat potential. The non-zero potential energy acts
as dark energy, causing the Universe to expand exponentially,
while quantum fluctuations of the inflaton field are magnified and
eventually, when the inflaton decays, are imprinted on the decay
products.

The large dark energy of the early Universe provides us with three
things: superluminal stretching of space needed to solve the
horizon problem; quantum fluctuations of scalar fields which may
seed density perturbations; and, once the primordial dark energy
decays, the origin of all matter. Scalar field inflation can model
all these three things but also has some obvious shortcomings; in
particular, it is totally unclear how, exactly, the Standard Model
(SM) degrees of freedom are produced. One is required to refer to
ad hoc couplings of the inflaton to some other fields, which then
couple to the SM fermions and bosons. Indeed, despite the apparent
success of simple inflationary models based on slowly evolving
scalar fields, there is no compelling theory of inflation.

Conventionally one relates the spectrum of density perturbations to
the properties of the inflaton potential. However, it has become clear that this is not a
necessary condition for a successful inflationary scenario. In curvaton
models \cite{Sloth,Lyth,Moroi} the primordial dark energy induces quantum fluctuations in an
effectively massless field whose energy density during inflation is negligibly small but
which later becomes dominant and then by decaying imprints the primordial perturbation on
matter fields.  Such a possibility has recently received a great deal of attention
\cite{Lyth1,Lyth3,Enqvist1,Enqvist2,Enqvist3,Postma,Kawasaki,Others}.
Because  during inflation the curvaton is massless at the scale
of the Hubble rate $H$, it will be subject to fluctuations which have a spectrum
that is almost identical to the spectrum of the inflaton field and therefore almost scale invariant.
However, as the curvaton is subdominant during
inflation, its perturbations will not be adiabatic but perturbations in the entropy density, i.e.
isocurvature perturbations.
When the curvaton decays, the isocurvature perturbations will be
converted to the usual adiabatic perturbations of the decay products,
which thus should ultimately contain also SM degrees
of freedom.

Litterature abounds with schemes of multifield inflation, but the curvaton models differ from
these in one important respect. It provides a logical separation between the field
responsible for density perturbations and for the geometrical stretching of space (such a
separation was already apparent in pre-Big Bang models,
see e.g.~\cite{Lidsey}). This separation opens the door to the possibility of
truly understanding the origin of all matter. Indeed, an exciting possibility could be
that the curvaton is just one of the fields of the Minimally Supersymmetric Standard Model (MSSM) \cite{Enqvist1,Enqvist2,Enqvist3,Kawasaki}
and that hence cosmology could be indirectly tested also by particle physics experiments.

\section{The curvaton paradigm}
During inflation, any effectively massless scalar field is subject
to fluctuations similar to the conventional inflaton with a
spectrum
\begin{equation}
{\cal P}^{1/2}(k)\simeq {H_*\over 2\pi}~,
\end{equation}
where $H_*$ is the value of the Hubble parameter during inflation
(at the horizon exit $k=aH$). By definition,  during inflation the
energy density of the inflaton must dominates. Therefore the
fluctuations in other scalar fields do not perturb space-time.
Such isocurvature perturbations may be converted to adiabatic
density perturbations if the subdominant scalar at some later
stage becomes dominant and then decays, as first pointed out in
\cite{Mollerach}. This idea was revived in the context of pre-Big
Bang models \cite{Sloth} and was then applied to conventional
inflation \cite{Lyth,Moroi} and dubbed as the curvaton mechanism.

Curvatons have been considered in the context of supersymmetric
theories \cite{Enqvist1,Enqvist2,Enqvist3,Kawasaki,susycurv}, and
both axions \cite{axioncurv} and pseudo-Nambu-Goldstone bosons
\cite{nambugoldst} have been suggested as curvaton candidates.
Curvaton dynamics, reheating and observational constraints have
been discussed in \cite{constraints}. Brane-inspired scenarios
have been coupled to the curvaton idea in
\cite{branescurv,piloriotto,emp}.

In its simples realisation, a curvaton model can be written as
\begin{equation}
V(\varphi,\phi)=V_{inf}(\varphi)+\frac 12 m^2\phi^2~,
\end{equation}
where $\varphi$ is the inflaton and $\phi$ the curvaton. $V_{inf}$
is the inflaton potential with $V_{inf}\simeq
H_*^2M_P^2\simeq\rho_{inf}$, where $M_P$ is the Planck mass. While
inflation lasts, the curvaton field remains fixed at some initial
value $\phi_*$ (in more complicated potentials, it may change
albeit typically very slowly). The curvaton's subdominance
requires $V_{inf}\gg m^2\phi_*^2$, and the curvaton is effectively
massless during inflation provided $m^2\ll H_*^2$. Then the
curvaton perturbation reads $\delta\phi_*\simeq H_*/2\pi$. The
perturbation will be gaussian if the mean value of the field is
larger than the perturbation itself, $\phi_*^2\gg H_*^2/4\pi^2$.

Once inflation is over, the inflaton field begins to oscillate
about the minimum of the potential and then decays. The curvaton
field starts to move away from $\phi_*$ and when $m^2\simeq H^2$,
the curvaton starts to oscillate. Oscillations behave effectively
as cold matter (in some background) and eventually overwhelm the energy density of the
inflaton decay products. The curvaton must decay before the onset
of nucleosynthesis so that the decay rate
$\Gamma_\phi>H_{nuc}\simeq 10^{-40}M_P^2$. A negligible curvature
perturbation due to inflaton means that curvaton models predict no
detectable primordial gravitational waves in the CMB.

The details depend on the actual form of the curvaton potential as
well as on the inflaton decay mechanism. The motion of the
curvaton field after inflation is particularly important as it can
affect the amplitude of the initial perturbation
($\delta\phi_*/\phi_*=\delta\phi/\phi$ only for a quadratic
potential). Curvaton dynamics has been discussed in \cite{Lyth3}
where it was found that if non-renormalizable terms dominate,
there exists an attractor solution which can erase the curvaton
perturbations. However, the conclusion depends somewhat on the
initial conditions.

\section{Flat directions in the MSSM}

An obvious candidate for a massless curvaton would be a flat direction
of the MSSM (for a review, see~\cite{Enqvist02}).
These are described in terms of order parameters, which are
combinations of squarks, sleptons and Higgses, which in the limit
of exact supersymmetry (SUSY) have a vanishing potential.
The existence of such flat directions is a consequence of gauge invariance
and supersymmetry. An example of an MSSM flat direction
is
\begin{equation}
\label{HLexample}
H_u=\frac1{\sqrt{2}}\left(\begin{array}{l}0\\ \phi\end{array}\right),~
L=\frac1{\sqrt{2}}\left(\begin{array}{l}\phi\\ 0\end{array}\right)~,
\end{equation}
where $H_u$ is the Higgs, $L$ a slepton and
$\phi$ is a complex field parameterizing the flat direction. All the other fields are
set to zero.
The MSSM flat directions have all been classified in \cite{gherghetta96}
and can be presented in terms of composite gauge invariant monomial operators\footnote{Flat directions can also be constructed from more than two
MSSM scalars, see \cite{askoJCAP}.}.

Flatness is lifted by SUSY breaking and by non-renormalizable
terms \cite{dine96,gherghetta96} of the form
$W=\lambda\Phi^n/nM^{n-3}$, where $M$ is the large cutoff scale,
usually taken to be the Planck scale, and $n\gtrsim 4$ is the
dimensionality of the non-renormalizable operator. The operators
of interest lifting the flatness turn out to have
dimensionalities $n=4,6,7,$ and 9.

There can also arise terms induced by the expansion of the
Universe which depend on the Hubble parameter $H$. The generic
form of the potential for an MSSM flat direction during inflation
is \cite{dine95,dine96}
\begin{equation}
\label{adpot0}
V(\phi)=(m^{2}_{\phi}-C H^2 ){|\phi |}^2 + \left[\left(a {\lambda}_n H + A_{\phi} \right){\lambda}_n {{\phi}^n \over n
M^{n-3}} + {\rm h.c.}\right]
+|\lambda|^2\frac{|\phi|^{2n-2}}{M^{2n-6}} \,.
\end{equation}
Here $\Phi=\phi e^{i \theta}/\sqrt{2}$.
The first term in Eq. (\ref{adpot0}) includes the Hubble-induced and low energy soft
mass terms. The second term includes the
Hubble-induced and low energy SUSY breaking $A$ terms. The Hubble-induced terms are
present only for F-term inflation; in the case of D-term inflation, they are
absent (for a discussion, see ~\cite{Enqvist02}). The last term is
the contribution from the non-renormalizable superpotential. The coefficients
$|C|,~a,~\lambda_{n}\sim {\cal O}(1)$, and the coupling
$\lambda \approx 1/(n-1)!$.

During inflation the MSSM scalar fields will fluctuate along the
flat directions  and, as long as these remain effectively
massless, form condensates. Which direction among the many
possible flat directions is chosen by the fluctuations is a matter
of chance. Note however that once one particular flat direction
starts to develop a condensate, its existence will generally
speaking lift the flatness of the other potential directions
(although there are some examples of directions which are
simultaneously flat, e.g. directions which involve only squarks
and only leptons). Because the observable Universe
originates from a single horizon patch, there is typically only
one MSSM condensate in our Universe.

\section{Flat directions as curvatons}

The $A$-terms in Eq. (\ref{adpot0}) are responsible for generating a global
charge for the condensate. This is important in the context of Affleck-Dine
baryogenesis \cite{Affleck:1984fy} and Q-balls \cite{qballs}, but for the present
considerations it suffices to focus on the amplitudes only. A generic form
of the flat direction potential contains a renormalizable mass term and a non-renormalizable
part and reads thus
\begin{eqnarray}
    V(\phi) & = &\frac{1}{2}m_{\phi}^2\phi^2 + V_{NR}, \\
    V_{NR} & = & \frac{\lambda^2 \phi^{2(n-1)}}{2^{n-1}M^{2(n-3)}},
    \label{potential}
\end{eqnarray}
where $m_{\phi}$ could also include a Hubble-induced
term. However, in that case the
amplitude of the fluctuations of the flat direction dies out
completely during inflation. For the curvaton mechanism to work
one must therefore require that the inflation model is such that
{\cal O($H$)} terms are not induced. One example is the SUSY
D-term inflation, which naturally leads to a vanishing Hubble-induced
mass term \cite{Kolda:1998kc}.
Another example invokes the so-called Heisenberg
symmetry on the K\"ahler manifold \cite{gaillard95}. In the latter
case, no mass term appears in the tree-level potential, although a harmlessly
small negative mass squared is induced at the
one-loop level. No-scale supergravity
theories belong to this class.

Let us simply assume that for one reason or another, during
inflation the flat direction does not receive any Hubble-induced
mass. Hence the mass term in Eq. (\ref{potential}) is of the order
of the SUSY breaking with $m_{\phi}\sim$ TeV $\ll H_*$. However,
this is not yet sufficient for the curvaton scenario to work. One
should also make sure that the inflaton fluctuations do not
contribute significantly to the adiabatic density perturbations.
This means that the Hubble parameter during inflation should obey
$H_* \sim \rho_{inf}^{1/2}M_P< 10^{14}$ GeV. (Needless to say, the
energy density of the flat direction should be negligible compared
to that of the inflaton, $\rho_{\phi} \ll \rho_{inf}$.) The
(isocurvature) fluctuation of the flat direction reads then
$\delta\phi \sim H_*/2\pi$, and if $\delta\phi/\phi_* \sim
H_*/\phi_* \sim 10^{-5}$, one obtains the right order of magnitude
for the density perturbation, provided there is no later damping
of the curvaton perturbation. Here $\phi_*$ is the amplitude
during inflation, and can be estimated by $V''(\phi_*) \sim
H_*^2$.

Simple analysis shows \cite{Enqvist1} that during inflation the curvaton is slow-rolling in the
nonrenormalizable potential $V_{NR}$. Thus, the Hubble
parameter and the amplitude of the field read,
respectively,
\begin{eqnarray}
    H_* & \sim & \lambda^{-\frac{1}{n-3}}\delta^{n-2 \over n-3}M_P,\\
    \phi_* & \sim & \lambda^{-\frac{1}{n-3}}\delta^{1 \over n-3}M_P,
\end{eqnarray}
where $\delta\equiv \delta\phi/\phi_* \sim H_*/\phi_*$.

After inflation, the inflaton field first oscillates and then
decays. The Hubble rate starts to decrease from its value $H_*$
during inflation and the MSSM curvaton moves towards the origin of
the field space and begins to oscillate when $H\sim m_\phi$. As a
consequence, its energy density relative to the inflaton decay
products starts to grow (assuming now for simplicity that the
inflaton has decayed before the curvaton oscillations begin). If
the decay products consist of (MS)SM radiation, there may arise
important thermal corrections in the effective curvaton potential.
In fact, it has been argued that the flat direction decays by
thermal scatterings before its domination and/or never dominates
the energy density of the Universe \cite{Postma, Enqvist2}.

These difficulties can be avoided by taking the inflaton sector to be hidden and
completely decoupled from the observable one \cite{Enqvist2,Enqvist3}.
Since the inflaton must be very weakly coupled to ordinary particles,
this seems rather a natural assumption which have in the conventional
inflationary picture has been precluded only because of the need to somehow generate
the SM degrees of freedom.
A hidden inflaton would decay into (light) particles in the
hidden  sector, and the inflaton decay products would not come into thermal
contact with the MSSM particles originating from the eventual curvaton decay.

Such a radical proposal may be quite welcome, because in spite of
the success of cosmic inflation in explaining the flatness and the
homogeneity of the Universe, there is not a single realistic
particle physics model which would give rise to an inflaton field
as a natural consequence of first principles. In almost all cases
the inflaton is a gauge singlet by construction and the contrived
slope of the inflaton potential is simply adjusted to fit the
observations. The coupling of such a singlet to the SM degrees of
freedom is usually set by hand. In this regard the MSSM flat
directions can play a significant role. A natural outcome of the
MSSM curvaton dynamics is the direct reheating of the Universe
with the MSSM degrees of freedom

\section{Constraints on the MSSM curvaton}

The MSSM curvaton should satisfy two conditions: 1) its energy
density should dominate the Universe at the time of its decay; 2)
the perturbations generated during inflation should not be damped
by the subsequent dynamics.

Let us begin by addressing the first issue. As already mentioned above, the oscillation of the
flat direction starts when $H \sim m_{\phi}\sim$ TeV, and the amplitude at
that time is  \cite{Enqvist2}
\begin{equation}
    \label{osc}
    \phi_{osc} \sim \left(\frac{m_{\phi}M^{n-3}}{\lambda}
    \right)^{1 \over n-2}.
\end{equation}
If the decay of the inflaton occurs earlier, at the onset of
curvaton oscillations the Universe will be dominated by hidden
radiation. The curvaton must nevertheless grow to dominate the
energy when it decays. Since the evolution of the energy density
of the flat direction is $\propto a^{-3} \propto H^{-3/2}$, one
can find  that the curvaton density equals the density of hidden
radiation $\rho_h \sim H_{EQ}^2$ when
\begin{equation}
    H_{EQ} \sim m_{\phi}\left(\frac{m_{\phi}M^{n-3}}
      {\lambda M_p^{n-2}}\right)^{4 \over n-2}.
\end{equation}
The curvaton should decay later than this so that the decay rate $\Gamma_{\phi} \sim f^2 m_{\phi}<H_{EQ}$, where $f$ is some Yukawa
(or gauge) coupling. Hence one obtains a constraint on coupling strength $f$. It turns out \cite{Enqvist2} that
the constraint is virtually identical also in the case when the inflaton decays after the curvaton oscillation
has already started, and that only the
$n \geq 7$ cases satisfy the decay-after-domination condition, with $n=7$ being marginal.

\begin{figure}
\includegraphics[width=80mm]{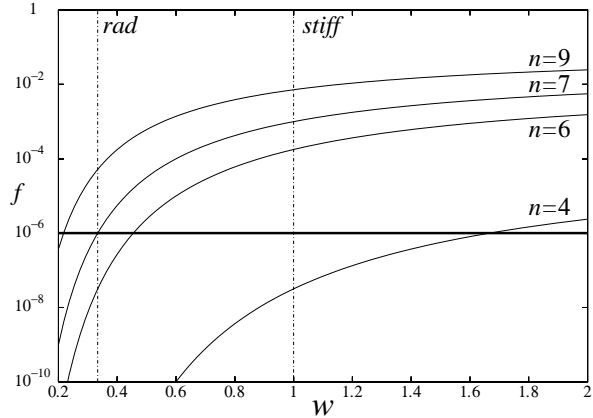}
\caption{\label{hid}
Constraint on the coupling for $n=4,6,7$, and 9. The allowed region is
below these lines and above $f=10^{-6}$, which is the smallest
coupling in MSSM. From \cite{Enqvist2}.}
\end{figure}

One may relax this condition if the hidden sector inflaton decays
not into radiation but into some fluid with a general equation of
state $p=w\rho$. Assuming the hidden fluid dominates the density
of the Universe at the time when the curvaton starts to oscillate,
one finds \cite{Enqvist2} the constraint
\begin{equation}
    f < \left[ \left(\frac{m_{\phi}}{\lambda M_p}
      \right)^{1 \over n-2}\right]^{\frac{1+w}{2w}}.
\end{equation}
This is depicted for $n=4,6,7,$ and 9 in Fig.~\ref{hid}. As one
can see, the $n=9$ direction is essentially the only usable for
the hidden radiation case, but even $n=6$ directions can be
available if the hidden sector fluid has a stiff equation of state
($w=1$). Notice that $n=4$ directions cannot dominate the Universe
and hence are completely ruled out as curvaton candidates.

Let us now turn to discuss the evolution of perturbations. In an
$m_{\phi}^2 \phi^2$ potential both the homogeneous
and (linear) perturbation parts obey the same type of equation so that the
ratio $\delta\phi/\phi$ remains a constant. However, the MSSM curvaton
is rolling in the non-renormalizable part of the potential. The
equations of motion for the superhorizon modes of the homogeneous and fluctuation
parts are written respectively as
\begin{eqnarray}
    \label{sr-homo}
    & & 3H\dot{\phi}+V'(\phi) = 0, \\
    \label{sr-fl}
    & & 3H\delta\dot{\phi}+V''(\phi) \delta\phi = 0.
\end{eqnarray}
Using Eq. (\ref{potential}) one then finds
\begin{equation}
    \frac{\delta\phi}{\phi} \sim
    \left(\frac{\delta\phi}{\phi}\right)_i
    \left(\frac{\phi}{\phi_i}\right)^{2(n-2)},
\end{equation}
where $i$ denotes the initial values. During inflation, there is
essentially no damping, but after inflation the field amplitude
decreases as $\phi \sim (H M^{n-3}/\lambda)^{1/(n-2)}$ as the
Hubble parameter changes from $H_*$ to $m_{\phi}$. Hence there is
a possibility for a large suppression of the curvaton perturbation
generated during inflation \cite{Lyth3}. It turns out
\cite{Enqvist2} that because of insufficient energy density and/or
damping of fluctuations, MSSM flat directions cannot act as
curvatons if $n\le 7$. In contrast, for the $n=9$ flat direction
the condensate will dominate the energy density of the
Universe before it decays, and if  initially $\phi \le 0.3M_P$,
which does not appear to be an unreasonable assumption, the
condensate perturbations will not be considerably damped during
inflaton oscillations. This result may be considered as a proof of
existence for an MSSM curvaton.

\section{Dumping inflaton energy out of this world}

With the advent of brane world scenarios, where the Universe is
regarded as a three dimensional hypersurface embedded in a higher
dimensional bulk, it has become possible to couple the inflaton to
degrees of freedom that are not visible in our Universe. (For a
review of brane world cosmology, see~\cite{Quevedo2}). Therefore
in the context of MSSM curvatons, which require the inflaton decay
products to reheat some hidden sector, it is of interest to ask
whether the whole of the primordial dark energy could be dumped
out of the visible Universe. The local density of the
inflaton decay products shoúld also be diluted; in other worlds, the
decay products should not remain in the vicinity of the brane.
This could be achieved by redshifting the decay products into
infinity of a warped 5th dimension \cite{emp}.

To this end, consider a simple Randall-Sundrum-type \cite{RS1}
scenario with a three dimensional hypersurface carrying MSSM
degrees of freedom. The MSSM brane is embedded in a $5$
dimensional space (the bulk) with a non-factorizable metric
\begin{equation}
\label{metric}
ds^2=e^{-2k|z|}(z)\eta_{\mu\nu}dx^{\mu}dx^{\nu}-dz^2\,,
\end{equation}
where $\eta_{\mu\nu}$ is the four dimensional Minkowski metric. We
take the  extra dimension to be  infinite. The  brane is located
at $z=0$. The action consists of
the 4D brane with tension $\sigma$ and a 5D bulk with a negative
cosmological constant which is related to the brane tension by
fine tuning. The warp factor present in the metric is $
k={4\pi}G_{5}\sigma/{3}$, where $G_{5}$ is the 5D Newton's
constant. The important point is that the geometry (\ref{metric}) has a particle horizon at
$z=\infty$: a particle that escapes from the brane and moves along
a geodesic travels from $z=0$ to $z=\infty$ in a finite proper
time $\tau=\pi/2k$.

Let us assume that the inflaton $\phi$ is a true
4D brane field, with a homogeneous distribution that dominates the
energy density in the early Universe on the observable brane and
gives rise to a period of inflation. Once inflation comes to an
end, the inflaton will decay, but instead of reheating the brane
degrees of freedom, let us assume that it couples to the bulk
scalar fields $\psi$ alone\footnote{For instance, by virtue of
some global quantum number carried by the inflaton and the bulk
fields but not by the MSSM fields.}, and hence decays into
Kaluza-Klein (KK) modes of the bulk degrees of freedom. This
happens through an effective coupling
\begin{equation}
 \sqrt{g(z)}\, h\, \phi(x)\,
 [\chi(0,m) \chi(0,m')]\, \psi_m(x)\, \psi_{m'}(x)\,;
\end{equation}
where $h$ is the coupling strength, $g(z)$ the metric and $\chi(0,m)$ are the ($z$ dependent) wave functions of a KK
modes of mass $m$. One may estimate the total decay rate as \cite{emp}
 \begin{equation}
   \Gamma_\phi = \int_{0}^{m_\phi}\int_{0}^{\sqrt{m_\phi^2 - m^2}}\,
            {dm\over k}{dm'\over k}\, h^2\,
        {[\chi(0,m) \chi(0,m')]^2 \over m_\phi}\,
     \approx
        {h^2\over 32} \left({m_\phi\over k}\right)^2\, m_\phi \,.
 \label{gamma}
 \end{equation}
The KK mass dependence of the effective couplings indicates that
the inflaton would preferably decay into the heavy modes $k\sim
M_{P}$, i.e. those with the largest momentum along the fifth
dimension. As the bulk modes carry momentum along the fifth
dimension, they would simply fly into the empty bulk and towards
infinity, taking the inflaton energy away from the brane. This process can
be thought of as a hot radiating plate cooling down by emitting
its energy into the cold surrounding space.

As the original energy density of the inflaton field is redshifted
away from the brane into the extra dimension, only the tail of the
density distribution would be felt by the brane. It has been
pointed out, however, that the energy emitted from the brane would
eventually collapse to a black hole at the end of the
5th dimension~\cite{March-Russell}. The presence of the black hole changes
the brane expansion by  introducing a new contribution to the
Friedmann equation which behaves as ${2 M_{BH}/ a^4}$, where  $a$
is the brane scale factor and $M_{BH}$ is a parameter interpreted
as the 5D "mass" of the black hole~\cite{visser}. This term acts
as a dark energy, which, provided that $M_{BH}$ is small, has a
subleading role in the early Universe.

There are some string theoretical motivations for a brane-world
scenario with warped infinite dimension involving $\bar D3$ and
$D3$ branes that annihilate at an adS throat, as discussed in
\cite{emp}. A similar string inflation model requiring no
slow-roll has been presented in \cite{piloriotto}.

\section{MSSM Higgses as curvatons}

If the inflaton reheats the bulk but not the brane, then MSSM
condensates do not decay because of thermal background and will
naturally dominate the energy density once the primordial dark
energy has been radiated away. In that
case several flat directions are good curvaton candidates,
including  \cite{Enqvist3} the simplest $n=4$, $H_uH_d$-direction
with
\begin{equation}
\label{example} H_u=\frac1{\sqrt{2}}\left(\begin{array}{l}0\\
\phi\end{array}\right)\,,~
H_d=\frac1{\sqrt{2}}\left(\begin{array}{l}\phi\\
0\end{array}\right)~\,.
\end{equation}
Only large amplitudes are relevant for cosmological purposes so
that one can ignore the $\mu$-term and the Higgs mass terms. Then
the effective potential for the $H_u H_d$ flat direction can be
written as
\begin{equation}
    V(\phi) = \lambda^2 \frac{|\Phi|^6}{M_p^2}\,,
\end{equation}
where $\lambda$ is a constant of ${\cal O}(1)$ which we set to
unity for simplicity. Writing  $
    V''(\phi_*) = \beta^2 H_*^2$ and $
   {H_*}/{\phi_*}=\delta
$ one requires that $\beta \ll 1$ and the perturbation $\delta
\sim 10^{-5}$. Thus one obtains
\begin{equation}
    \phi_* \sim \beta\delta M_P, \qquad
    H_* \sim \beta \delta^2 M_P\,.
\end{equation}
One then finds the scale of inflationary dark energy to be
$V_{inf}^{1/4} \sim (H_*M_P)^{1/2} \sim \beta^{1/2}\delta~M_P$.

Once the condensate decays, it will reheat the universe. The
maximum temperature is $T_{max} \sim [V(\phi_*)]^{1/4}$. After
inflation the amplitude of the flat direction is very large so one
expects both fermionic and bosonic preheating \cite{preheating}.
(Preheating in the context of MSSM flat direction has been
discussed in \cite{postmapr}). The $q$-parameter in this case is
given by \cite{Enqvist3}
\begin{equation}
    q \sim \frac{f^2\phi_*^2}{\omega^2} \sim f^2
    \left(\frac{M_p}{\phi_*}\right)^2 \gg 1\,,
\end{equation}
where $\omega = \sqrt{V''(\phi_*)}$. Once MSSM fermions and bosons
are produced, the energy density of (non-thermal) radiation
evolves as $\rho_{rad}=\rho_f+\rho_b \propto a^{-4}$, while the
residual oscillation of the flat direction decreases as
$\rho_{\phi} \propto a^{-9/2}$, where $a(t)$ is the scale factor
of the universe. After the mass term in the effective potential of
the flat direction dominates,
$\rho_{\phi} \propto a^{-3}$. The equality of the energy densities
occurs when $\phi=\phi_{eq} \sim (\beta\delta)^{-1} m$, where $m$
is ${\cal O} (100)$ GeV. If the decay of the flat direction takes
place before this time, the contribution to the total radiation
density is subdominant. This happens if
\begin{equation}
    f \gtrsim \sqrt{8\pi}\left(\frac{\phi_{eq}}{M_p}\right)^{1\over2}
    \sim \sqrt{8\pi}(\beta\delta)^{-\frac{1}{2}}
    \left(\frac{m}{M_p}\right)^{1\over2}\,.
\end{equation}
For $\beta=0.1$ and $\delta=10^{-5}$, RHS reads $\sim 10^{-4}$.
Since the Higgs has couplings much larger than $10^{-4}$ (e.g.
gauge couplings), the conclusion is that the $H_u H_d$ flat direction decays well before
the equality time. Hence the amount of radiation is totally
determined by the preheating era. One finds \cite{Enqvist3} that
the reheat temperature $
    T_{RH} \lesssim T_{max} \sim 10^9
    \left(\frac{\beta}{0.1}\right)^{3\over 2}
    \left(\frac{\delta}{10^{-5}}\right)^{3\over 2} {\rm GeV},
$ so that the gravitino problem~\cite{Ellis} is automatically
avoided (for a generic $n=6$ ($n=7$) direction the maximum reheat
temperature would be $10^{13}~(10^{14})$ GeV).

The spectral index of the CMB temperature perturbations can be
evaluated as \cite{Lyth}
\begin{equation}
\label{nspectr}
    n_s-1 = 2 \frac{\dot{H_*}}{H_*^2}
    + \frac{2}{3}\frac{V''(\phi_*)}{H_*^2}.
\end{equation}
For $H_uH_d$ the change of the Hubble parameter is negligible and
one finds \cite{Enqvist3} $n_s-1 \approx 0.007$ for $\beta=0.1$.
Note that there is a dependence on the Higgs potential, which in
principle could be determined in the laboratory by extracting the
relevant N-point scattering amplitudes from the data.

\section{Discussion}

If inflation takes place in a hidden sector, the $n=9$ flat direction $QuQue$
can act
as a curvaton and produce the required spectrum of density perturbation. Since
the MSSM curvaton consists of squark and slepton fields, its decay will naturally
give rise to both baryons and cold dark matter in the form of neutralinos.
A possible complication is that instead of decaying directly
the condensate may first fragment into Q-balls \cite{qballs,Enqvist02}.
This is also the generic behaviour of the $n=9$ flat direction in, say,
a no-scale model (but does not happen for the
3rd generation $QuQ_3ue$ direction if ${\rm tan}\beta\le 3$) \cite{Enqvist2}.  Another complication could be the fact that since the
MSSM curvaton is a complex field, its phase will also be subject to perturbations. These
correspond to isocurvature fluctuations of the global charges $B$ and/or $L$ and may give rise
to a baryonic isocurvature perturbation which is too high \cite{Kawasaki}. However,
$QuQue$ happens to have a vanishing $A$-term and is therefore safe from this constraint \cite{Enqvist2}.

The implicit assumption here has been that the inflaton decay
products determine the early evolution of the Universe before the
curvaton becomes dominant. As discussed here, with the advent of extra
dimensional theories such as brane worlds, this is no longer a
logical necessity. If our Universe is a brane-like object
embedded in a higher dimensional bulk, it is possible that the
primordial dark inflaton energy could be radiated out of the brane
into the bulk. The analogue would be a a hot plate emitting its
energy into the surrounding cold space. It has been argued that
such a situation could be found within the brane world scenarios
with a warped, infinite extra dimension \cite{emp}. In such a case
the inflaton decay products would move rapidly away from the
vicinity of the brane and redshift into the infinity of the 5th
dimension. Then almost any MSSM flat direction could be the
curvaton, including the simplest one based on the Higgses $H_u$
and $H_d$ \cite{Enqvist3}. One would then find a spectral index very close to
1 but with a weak dependence on the Higgs potential.

It is an open problem whether dumping of the inflaton energy into extra dimensions
can be realised in a string theoretical context.
Meanwhile, the MSSM curvaton remains an interesting possibility for the origin of all
matter and density perturbations that could, at least in principle, be testable
also in the laboratory.

\section*{Acknowledgments}

I am grateful to Asko Jokinen, Shinta Kasuya and Anupam Mazumdar for many useful
discussions. This work is partially supported by the Academy of Finland
grants 75065 and 205800.



\end{document}